\documentclass[12pt,journal,onecolumn,draftclsnofoot]{IEEEtran}

\def\argmax{\mathop{\rm arg\,max}}

\usepackage{graphicx,epsfig}
\usepackage{subfigure}
\usepackage{cite}
\usepackage{amsmath}
\usepackage{amssymb}
\usepackage{amsfonts,helvet}
\usepackage{footmisc}
\usepackage[alwaysadjust]{paralist}
\usepackage{url}

\newtheorem{defn}{Definition}

\newtheorem{lemma}{Lemma}

\newtheorem{proposition}{Proposition}
\newtheorem{remark}{Remark}

\newtheorem{assumption}{\textbf{AS}}

\def\cB{\mbox{$\mathcal{B}$}}

\def\cH{\mbox{$\mathcal{H}$}}

\newcommand{\abs}[1]{\lvert #1\rvert}
\newcommand{\vectornorm}[1]{\left|\left|#1\right|\right|}

\begin{document}
\title{Spectrum Allocation in Two-Tier Networks}



\author{\authorblockN{Vikram Chandrasekhar and Jeffrey G. Andrews} \\
\thanks{This research has been supported by Texas Instruments.
The authors are with the Wireless Networking
and Communications Group, Dept. of Electrical and Computer
Engineering at the University of Texas at Austin, TX 78712-1157.
(email:cvikram@mail.utexas.edu, jandrews@ece.utexas.edu), Date: \today.}}


\maketitle

\begin{abstract}
Two-tier networks, comprising a conventional cellular network
overlaid with shorter range hotspots (e.g. femtocells, distributed
antennas, or wired relays), offer an economically viable way to
improve cellular system capacity. The capacity-limiting factor in
such networks is interference.  The cross-tier interference between
macrocells and femtocells can suffocate the capacity due to the
near-far problem, so in practice hotspots should use a different
frequency channel than the potentially nearby high-power macrocell
users.  Centralized or coordinated frequency planning, which is
difficult and inefficient even in conventional cellular networks, is
all but impossible in a two-tier network.  This paper proposes and
analyzes an optimum decentralized spectrum allocation policy for
two-tier networks that employ frequency division multiple access
(including OFDMA).  The proposed allocation is optimal in terms of
Area Spectral Efficiency (ASE), and is subjected to a sensible
Quality of Service (QoS) requirement, which guarantees that
both macrocell and femtocell users attain at least a prescribed data rate.
Results show the dependence of this allocation on the QoS requirement,
hotspot density and the co-channel interference from the macrocell and surrounding femtocells.
Design interpretations of this result are provided.
\end{abstract}

\section{Introduction}
Wireless operators are in the process of augmenting the macrocellular network with supplemental infrastructure such as microcells\cite{Shapira1994,Ganz1997,I1993}, distributed antennas\cite{Saleh1987} and relays. An alternative with lower upfront costs is to improve indoor coverage and capacity using the concept of \emph{end-consumer} installed femtocells or home base stations\cite{Doufexi2003}. A femtocell serves as a low power, short range data access point that provides high quality in-building coverage to home users, while backhauling their traffic over the IP network. The femtocell radio range ($10-50$ meters) is much smaller than the macrocell radius ($300-2000$ meters) \cite{Ganz1997}. Users transmitting to femtocells experience superior indoor signal reception and lower their transmit power, consequently prolonging battery life. The implication is that femtocells provide higher spatial reuse and cause less cochannel interference (CCI) to other users.
The principal arguments in favor of femtocells are summarized in \cite{ChandrasekharMag2008}.

In a shared spectrum two-tier network  (universal frequency reuse), recent research \cite{ChandrasekharPC2008} has shown that near-far effects arising from cross-tier interference create a fundamental trade-off: Assuming all femtocell users seek a common SIR, the product of all Pareto-optimal macrocell and femtocell SIR targets is a constant dependent only on the intra- and cross-tier channel powers, irrespective of the power control strategy. Perron-Frobenius theory\cite{Horn1985} shows that any combination of per-tier target SIRs whose product violates the above constraint causes the spectral radius of the normalized channel power matrix to exceed unity; hence, there may not be a feasible power allocation for all users. The worst-case scenario arise either when a high powered macrocell user on the cell edge causes interference to nearby femtocells, or when cell interior femtocell users cause unacceptable interference to the macrocell BS. The papers \cite{Hoven2005,Sahai2006,ChandrasekharPC2008,Chandrasekhar2007} have suggested addressing this near-far problem by either \begin{inparaenum} \item forcing femtocell users to decrease their target data rates, or \item employing interference avoidance (e.g. randomized time hopping) among both macrocell and femtocell users in order to ``avoid" transmitting in the same interval. \end{inparaenum} Both approaches have drawbacks because they require either reducing SIR targets at femtocells, or accommodating femtocells by altering the transmission scheme at an existing macrocell in order to ensure tolerable performance. The subject of this paper is to motivate and propose a spectrum partitioning strategy in a two-tier deployment.

\subsection{The Return of FDMA}
Frequency division multiple access (FDMA)'s  resurgence in emerging OFDMA wireless standards such as the 3GPP's LTE, WiMAX and 3GPP2's UMB enable the macrocell to perform flexible rate assignment\cite{Andrews} across frequency sub bands to users and provide interference management by fractional frequency reuse. In femtocell deployments, due to reasons of scalability, security and limited availability of backhaul bandwidth, it is reasonable to assume the absence of coordination between femtocells and the central macrocell. Further, femtocells are placed opportunistically or randomly by end users. Therefore, conventional frequency planning strategies will be very difficult in a two-tier network.

Assigning orthogonal spectrum resources between the central macrocell and femtocell BSs eliminates cross-tier interference. This motivates the orthogonal access spectrum allocation strategy proposed in this paper. Next, to avoid persistent collisions with neighboring femtocells in their allotted spectrum, this paper proposes that each femtocell accesses a random subset of the candidate frequency subchannels, wherein each subchannel is accessed with equal probability. We term this spectrum access strategy as F-ALOHA (Frequency ALOHA)\footnote{Slotted ALOHA, by convention, implies that the slots are in the time domain}. We motivate F-ALOHA for three reasons. First, F-ALOHA avoids transmission delays and increased RF sensitivity requirements for sensing frequency subchannels in the presence of channel fading\cite{Sahai2006}. Next, F-ALOHA provides randomized interference avoidance, since neighboring femtocells are unlikely to consistently access identical frequency subchannels. Finally, such a transmission strategy offers a decentralized spectrum access by femtocells--eliminating backhaul communication cost between femtocells over the internet--and low complexity.

\textbf{F-ALOHA Spectrum Access:} If a femtocell transmits over all its allotted subchannels, it may cause excessive interference to surrounding femtocells; conversely, accessing only a few subchannels could result in a poor spatial reuse. With F-ALOHA, there should be an optimal fraction of spectrum access for each femtocell in order to maximize the spatial reuse of spectrum, or in effect the net number of simultaneous transmissions per unit area\cite{Baccelli2006,Weber2007}. The spatial reuse is readily expressible using the \emph{Area Spectral Efficiency} (ASE) in $\textrm{b/s/Hz/m}^2$\cite{Alouini1999}, which is defined as the network-wide spatially averaged throughput per frequency subchannel divided by the product of the subchannel bandwidth and the area over which the transmissions take place. Based on the stated reasons, we assume \begin{inparaenum}[(1)] \item downlink transmissions from the macrocell and femtocells are frequency orthogonal and \item femtocells transmit using F-ALOHA, \end{inparaenum}and pose the following questions:
\begin{itemize}
\item What is the expected subchannel throughput inside the macrocell [resp. femtocell], as a function of interference from neighboring macrocells [resp. femtocells], and terrestrial propagation parameters such as path-loss exponent and lognormal shadowing?
\item Given the expected subchannel throughput, the average number of femtocells per cell-site and the number of users associated with each BS, how should the bandwidth be partitioned between tiers in order to satisfy a Quality of Service (QoS) requirement in each tier?
\item With this spectrum allocation strategy, how much improvement in network-wide ASE does opportunistic channel aware macrocell scheduling offer relative to channel blind scheduling?
\end{itemize}

\subsection{Related Work}
Existing research on hierarchical cellular systems has mainly focused on channel assignment in two-tier macrocell/microcell radio systems\cite{Kishore2005,Shen2004,Yeung1996}. In the context of this paper, a microcell has a much larger radio range (100-500 m) than a femtocell, and generally implies centralized deployment, i.e. by the service-provider. These works typically either assume only a single microcell, or regularly spaced multiple microcells. Such an assumption may be impractical in femtocell networks because of the variations in their placement from one cell site to the next.

We would also like to clarify the differing objectives between a femtocell and microcell overlay. A microcell overlay allows the operator to handoff and load balance users between each tier in response to changing traffic conditions\cite{ChandrasekharMag2008}. So, the operator can preferentially assign users (e.g high data rate users) by converting microcells into data access points\cite{Kishore2005,Shen2004}. In contrast, femtocells are consumer installed and the traffic requirements at femtocells are user determined without any operator influence. In heterogeneous networks, \cite{Niyato2005,Yang2007} suggest providing vertical handoffs (admission control) and QoS-aware routing to switch traffic between cellular and \emph{ad hoc} technologies. However, these approaches may not be viable in two-tier femtocell networks because of the lack of coordination between macrocell and femtocell BSs; instead, decentralized strategies for interference management and load balancing may be preferred.

The problem considered in this paper is related to Yeung and Nanda \cite{Yeung1996}, who propose frequency partitioning in a microcell/macrocell system based on mobile speeds and the loading of users in each cell. Similar dynamic channel allocation schemes have been proposed in \cite{Lagrange1997} and \cite{Furukawa1994}. Their frequency partitioning is derived based on choosing handoff velocity thresholds and maximizing the overall system capacity, subject to per-tier blocking probability constraints that ignore co-channel interference (CCI). In contrast, our work determines the spectrum allocation which maximizes the system-wide ASE considering interference from neighboring BSs, path loss and prevailing channel conditions.

In decentralized networks, Jindal \emph{et al.}\cite{Jindal2007} have derived the optimal number of frequency subchannels $F$ for a frequency-hopped ad hoc network assuming a fixed data rate requirement per transmitter-receiver pair and a target outage probability. Our work, in contrast, assumes a fixed $F$ (as is the case with OFDMA), allowing for multiple subchannels to be accessed by each BS, with variable rate transmission per accessed subchannel. In a hybrid network composed of ad-hoc nodes and BS infrastructure nodes, Zemlianov and de Veciana\cite{Zemlianov2005} and Liu \emph{et al.}\cite{Liu2003} have derived asymptotic scaling laws relating to the per user throughput as the number of infrastructure nodes increase.  In addition to the hierarchical nature of our model, the main difference is that this paper assumes single-hop communication. In WLAN networks, Bahl \emph{et al.}\cite{Bahl2007} have proposed variable center frequencies and variable channel width per access point for improving the spectrum utilization and fairness for heavily loaded access points (APs). Finally, game-theoretic approaches have been recently investigated for both non-cooperative \cite{Etkin2007,grokop-2008} and cooperative \cite{Suris2007} spectrum sharing in decentralized networks.

\subsection{Contributions}
This paper employs a stochastic geometry framework for modeling the random spatial distribution of femtocells. Hotspot locations are likely to vary from one cell site to another, and be opportunistic rather than planned, so an analysis that embraces instead of neglecting randomness should provide more accurate results and more plausible insights. Towards this end, femtocells are assumed to be scattered according to a Spatial Poisson Point Process (SPPP)\cite{Kingman}, and inter-femtocell interference is modeled as Poisson Shot-noise\cite{Lowen1990}. This model has been used extensively in prior work \cite{Chan2001,Baccelli2001,Ridolfi2006}, and its validity has been confirmed in empirical studies\cite{Baccelli1997}.

The proposed spectrum allocation maximizes the network-wide ASE in a two-tier network, assuming each macrocell transmits to a single user per frequency subchannel, while femtocells access spectrum using F-ALOHA. The allocation is determined in two steps. First, the per-tier ASEs are quantified based on the propagation environment and neighboring cell interference. With an increasing number of macrocell users, the ASE of the macrocell is either fixed (for a channel blind scheduler) or increasing (by opportunistic scheduling). For the femtocell network, the ASEs are derived based on the optimal spectrum access using F-ALOHA. Next, the optimal allocation is determined as one that maximizes the weighted mean of the per-tier ASEs--the weights are given by the fraction of spectrum accessed by each tier. The three contributions of this paper are as follows.
\begin{asparadesc}
\item[Expected per-tier throughput.]  The expected per-tier throughput is derived for the macrocell and femtocell users accounting for interference from neighboring cells. The maximum ASE of the femtocell network is shown to be \emph{unchanged} with addition of hotspots beyond a threshold. At low femtocell densities, the highest femtocell ASEs are attained when each femtocell accesses the entire available spectrum. In higher densities, femtocells should use a decreasing fraction of the spectrum; e.g. with an average of $100$ femtocells in each cell site, each femtocell should access $30 \%$ of the available spectrum.
\item[Spectrum allocation with Quality of Service (QoS).] The proposed spectrum allocation maximizes the spatial reuse in a two-tier network, subject to a network-wide QoS requirement, which guarantees a minimum expected throughput per-user. Differing QoS constraints produce markedly different spectrum allocations due to the competing spatial coverage scales in each tier. Notably, a QoS requiring equal per-user throughputs in each tier means assigning greater than $90\%$ of spectrum to the macrocell. Conversely, an even division of spectrum occurs when the QoS constraints favor femtocells to provide significantly higher data rates.
\item[Scheduling and Spectrum Requirements.] Gains generated by channel aware macrocell scheduling permeate to femtocells, resulting in a significant spectrum reduction with the proposed allocation. With an average of $50$ femtocells/cell site and target per-tier data rates of $0.1$ Mbps/macrocell user and $10$ Mbps/hotspot user, a channel aware macrocell scheduler provides nearly $50\%$ reduction in necessary spectrum compared to a channel blind scheduler. Finally, with increasing number of hotspot users, the spectrum requirements in a two-tier network show two extremes. One is a low interference scenario where addition of hotspots provides increased spatial reuse, ensuring that the necessary spectrum is unchanged up to $110$ femtocells/cell site. In a high interference scenario however, the ensuing co-channel interference may necessitate a linear increase in required spectrum with hotspot density.
\end{asparadesc}

\section{System Model}
The setup consists of a hexagonal region $\cH$ of radius $R_c$ with a central macrocell BS $C$ providing coverage area $|\cH|=\frac{3\sqrt{3}}{2} R_c^2$, which is surrounded by two rings of interfering macrocells. The macrocellular network is overlaid with femtocell hotspots of radius $R_f$, which are randomly distributed on $\mathbb{R}^2$ according to a homogeneous SPPP $\Omega_f$ with intensity $\lambda_f$ \cite{Kingman}. The mean number of femtocells per cell site is readily obtained as $N_f=\lambda_f |\cH|$. Macrocell users are assumed to be uniformly distributed inside each cell site. Femtocells are assumed to provide ``closed access" to licensed indoor users who fall within the radio range $R_f$ of their respective home BSs\footnote{The closed access scheme permits unlicensed users within the femtocell radio range to communicate with the macrocell.}. Let $U=U_c+N_f U_f$ denote the average number of users in each cell site. These $U$ users are distributed into $U_c$ uniformly distributed tier 1 mobile outdoor users and $U_f$ users per femtocell hotspot.

\subsection{Per-Tier Spectrum Access}
The available spectrum comprises $F$ frequency subchannels each with bandwidth $W$ Hz. We wish to determine the optimal partitioning $(F_c,F_f)$, where $F_c$ subchannels are available for macrocell transmissions and $F_f=F-F_c$ subchannels are available for femtocell transmissions. Denote $\rho=F_c/F$ as the \emph{fraction of spectrum} assigned to the macrocell BS with the following key assumptions:
\begin{assumption}
Each femtocell schedules its users in a round-robin (RR) fashion. The macrocell schedules its users according to either a channel blind RR or a channel aware proportional fair (PF) scheduler.
\end{assumption}
\begin{assumption}
The fraction $\rho$ takes a continuum of values in $[0,1]$. For simplicity, the paper does not quantize $\rho$ for having an integer number of frequency subchannels. Consequently, $F_c = [\rho F]$ and $F_f = F-F_c$, where we use $[x]$ to denote the integer part of a number $x$.
\end{assumption}
Assumption $1$ makes it clear that the long term expected throughput per indoor user equals $1/U_f$th of the average femtocell throughput. The long term expected macrocell throughput is assumed to be equally divided among the $U_c$ outdoor users with RR and PF scheduling at the macrocell. With a PF scheduler, this assumption is reasonable considering mobility, which ensures that all users receive an identical average Signal to Interference Ratio (SIR) over the long term.

If each femtocell transmits over exactly $k$ frequency subchannels among their allotted $F_f$ subchannels, the net portion of accessed spectrum per femtocell equals $\rho_f(1-\rho)$ where $\rho_f \triangleq k/F_f$. If femtocells choose their frequency subchannels independently and with equal probability, F-ALOHA effectively ``thins" the mean number of interfering femtocells in each frequency subchannel. The probability $p$ of a femtocell selecting a given frequency subchannel for transmission is given as:
\begin{align}
\label{FemtALOHASelnProb}
p=\frac{\binom{F_f}{k}-\binom{F_f-1}{k}}{\binom{F_f}{k}}=\frac{k}{F_f}=\rho_f
\end{align}
Consequently, the set of interfering femtocells per frequency subchannel is a \emph{marked SPPP} \cite{Kingman} $\Lambda_f$ with intensity $\lambda_f \rho_f=\lambda_f k/F_f$. When $\rho_f=1$, all femtocells in $\Omega_f$ access the entire spectrum but mutually interfere in all subchannels. For $\rho_f \ll 1$, femtocells transmit in a small region of spectrum and avoid causing mutual interference. This strategy provides a higher spectral efficiency over each frequency subchannel, but incurs reduced spectrum utilization because femtocells do not transmit over the entire available spectrum.

\subsection{Channel Model and Variable Rate Transmission}
The downlink channel between each BS and its users is composed of a fixed distance dependent path loss, a slowly varying component modeled by lognormal shadowing and Rayleigh fast fading with unit average power. For simplicity, thermal noise is neglected at the receiver since cellular systems, by nature, are interference limited.
\begin{assumption}
\label{AS:SIR}
Each user is assumed to track their SIR in each subchannel and feedback the instantaneous rate to their BS with zero delay. Further, the channel can support the requested rate as determined by the scheduled user with probability 1. Although we acknowledge that imperfect feedback and/or channel estimation has a potentially big impact on system capacity, this paper does not account for these effects for sake of analytical tractability.
\end{assumption}
\begin{assumption}
\label{AS:TXPWR}
BSs assign equal transmission powers to all subchannels.
\end{assumption}
Each BS assigns rate adaptively based on the received SIR per user. Let $G$ denote the Shannon Gap with variable rate M-QAM transmission \cite{Goldsmith}. Assume an instantaneous transmission rate of $b_i$ bps/Hz if the instantaneous SIR lies in $[\Gamma_i, \Gamma_{i+1})$. Using adaptive modulation with $L$ discrete rates, the instantaneous rate $Wb$ in a $W$ Hz wide subchannel is chosen as:
\begin{align}
\label{Eq:InstRate}
b&=b_i, \textrm{when }  \mathrm{SIR} \in [\Gamma_i,\Gamma_{i+1}), 1 \leq i \leq L \\
b_i&=\log_2 \left (1+\frac{\Gamma_i}{G} \right ) \textrm{bps/Hz}
\end{align}
Assuming \emph{identical statistics} over all frequency subchannels, the long term expected throughput (in b/s/Hz) per macrocell/femtocell in each subchannel is given as:
\begin{align}
\label{Eq:MeanRate}
T=\sum_{l=1}^{L-1} l \cdot \Pr[\Gamma_l \leq \mathrm{SIR} < \Gamma_{l+1}]+L \cdot \Pr[\mathrm{SIR} \geq \Gamma_L]
\end{align}
The expected throughput provided by each macrocell [resp. femtocell] is obtained multiplying the expected throughput in \eqref{Eq:MeanRate} by their respective spectrum allocation $\rho$ [resp. $\rho_f(1-\rho)$].

\section{Spectrum Allocation and Per-Tier Expected Throughputs}
Let spectrum $WF$ be partitioned such that the macrocell BS transmits over a portion $\rho$, while femtocell BSs transmit over the remaining $1-\rho$ fraction of the spectrum. Let $T_c(\rho,U_c)$ be the long term throughput (in b/s/Hz) in each subchannel provided by the macrocell\footnote{The use of $U_c$ and $\rho$ within parenthesis is to account for a macrocell scheduler which can provide diversity gains by scheduling users according to their channel variations\cite{Viswanath2002}}. Obtaining $T_c$ requires calculating the average rate per subchannel in \eqref{Eq:MeanRate} after spatially averaging the SIR over all locations, and accounting for the interference from two rings of transmitting macrocells.

Let each femtocell access a portion $\rho_f$ of its allotted spectrum using F-ALOHA, servicing its users in a RR schedule. Define $T_f(\rho_f \lambda_f)$ as the expected femtocell throughput in each frequency subchannel, which is determined by the intensity $\rho_f \lambda_f$ of the marked SPPP $\Lambda_f$. With universal frequency reuse across all macrocells, the area spectral efficiency (ASE) of the macrocell (resp. femtocell) network is given as:
\begin{align}
\label{Eq:ASEMacroFemto}
\textrm{ASE}_c=\frac{T_c(\rho,U_c)}{|\cH|}, \quad
\textrm{ASE}_f=\frac{N_f \rho_f T_f(\rho_f \lambda_f)}{|\cH|}
\end{align}
The factor $N_f \rho_f$ represents the mean number of transmitting femtocells in each subchannel. With bandwidth $W$, the per-tier throughputs (in b/s) per subchannel can be calculated by multiplying the ASEs in \eqref{Eq:ASEMacroFemto} by $W|\cH|$. The network-wide ASE is therefore given as:
\begin{align}
\label{Eq:NetworkASE}
\textrm{ASE}&=\rho \textrm{ASE}_c+(1-\rho)\textrm{ASE}_f \notag \\
            &= \frac{1}{|\cH|}[\rho T_c(\rho,U_c)+(1-\rho)N_f \rho_f  T_f(\rho_f \lambda_f)]
\end{align}
The expected network throughput (in b/s) over the $WF$ wide spectrum is obtained by multiplying \eqref{Eq:NetworkASE} by $WF|\cH|$. Before determining the spectrum allocation, we first stipulate a QoS requirement $\eta$, which ensures that users in either tier are guaranteed a \emph{minimum} expected throughput. By implication, $\eta$ also regulates the maximum amount of spectrum that any tier can receive.
\begin{defn}
The QoS parameter $\eta$ guarantees that the expected throughput per user in one tier is \emph{at least} $\eta/(1-\eta)$ w.r.t the other tier.
\end{defn}
Choosing different $\eta$ enables assigning different priorities (QoS) to one tier relative to the other. For example, setting $\eta=0.5$ ensures that users in both tiers obtain identical expected rates. On the other hand, decreasing $\eta$ favors assigning greater spectrum to the tier providing a higher expected throughput per active user.

Given a total available spectrum of $1$ Hz, the problem is to determine the optimal spectrum allocation $\rho$ over all possible spectrum partitioning strategies $\omega \in [0,1]$ between the macrocell and femtocells. The proposed spectrum allocation maximizes the network-wide ASE with a QoS constraint $\eta$ on the minimum expected per-tier throughput/user, as shown below:
\begin{align}
\label{Eq:TwoTierOptASE}
\rho&= \frac{1}{|\cH|} \arg \max_{0 \leq \omega \leq 1} \omega T_c(\omega,U_c)+(1-\omega) N_f \rho_f T_f(\rho_f \lambda_f) \\
\label{Eq:TwoTierOptASEConstraint}
&\textrm{subject to } \min \left \lbrace T_{c,u}(\omega), T_{f,u}(\omega) \right \rbrace \geq \eta(T_{c,u}(\omega)+T_{f,u}(\omega) ) \\
&\textrm{where } T_{c,u}(\omega) \triangleq \frac{\omega T_c(\omega,U_c)}{U_c} \textrm{ and } T_{f,u}(\omega) \triangleq \frac{(1-\omega)\rho_f T_f(\rho_f \lambda_f)}{U_f} \notag
\end{align}
Here $T_{c,u}(\omega)$ and $T_{f,u}(\omega)$ are the \emph{expected throughputs for a macrocell and femtocell user} respectively.
Whenever the average subchannel throughput $T_c(\omega,U_c)$ is \emph{independent} of the spectrum $\omega$ assigned to the macrocell, the objective function in \eqref{Eq:TwoTierOptASE} is an affine function w.r.t $\omega$. The following proposition derives the optimizing $\rho$ considering that the maximum is attained at the extremal points of the constraint set:
\begin{proposition}
\label{Pr:TwoTierOptASE}
If the expected macrocell throughput per subchannel is independent of the total spectrum allocated to the macrocell $\omega$, i.e. $T_c(\omega,U_c)=T_c(U_c) \textrm{ } \forall \omega \in [0,1]$, the optimizing $\rho$ in \eqref{Eq:TwoTierOptASE} satisfies the QoS constraint with equality, belonging to a set with two candidate spectrum allocation assignments given as:
\begin{align}
\label{Eq:ASEOptRho}
\rho^{\ast} &\in \lbrace x,1-x \rbrace, x \triangleq \Biggl \lbrack 1+\frac{1-\eta}{\eta}\frac{T_c(U_c)}{U_c}\frac{U_f}{\rho_f T_f(\rho_f \lambda_f)} \Biggr \rbrack^{-1}
\end{align}
\end{proposition}
\vspace{5 mm}
\begin{proof}
Since $T_c(\omega,U_c)=T_c(U_c) \textrm{ } \forall \omega \in [0,1]$, the optimization problem in \eqref{Eq:TwoTierOptASE} is to determine the optimal $\rho$ which maximizes a convex combination of $T_c$ and $N_f \rho_f T_f(\rho_f \lambda_f)$ with a linear constraint. Consequently, the argument maximizer is located at the extremal points of the constraint set \eqref{Eq:TwoTierOptASEConstraint}. Solving for the $\rho$ which satisfies \eqref{Eq:TwoTierOptASEConstraint} with equality yields \eqref{Eq:ASEOptRho}.
\end{proof}
\begin{remark} Without a QoS requirement (allowing $\eta \rightarrow 0$), the objective function in \eqref{Eq:TwoTierOptASE} is a convex combination of the macrocell and femtocell throughputs which is maximized at the extreme points $\rho \in \lbrace 0,1 \rbrace$. Such a partitioning is clearly unfair since it results in a greedy allocation of the entire spectrum to one tier.
\end{remark}
For a generic macrocell scheduler--when Proposition \ref{Pr:TwoTierOptASE} may be inapplicable-- \eqref{Eq:TwoTierOptASE} is a one dimensional optimization problem that can be solved efficiently for a given $\eta$ using a numerical search.

\subsection{Macrocell Throughput: RR Scheduling}
Assuming that the central macrocell BS $C_0$ in the hexagonal region $\cH$ is placed at the origin, the normalized positions of the interfering BSs $C_k \in \cB, k=1 \dots 18$ are represented in polar form $\mathbf{b_k}, k \in \cB$ using MATLAB notation as:
\begin{align}
\mathbf{b_k}/R_c \in \Big \lbrace \sqrt{3}e^{i(\pi/6+[0:5]\pi/3)}\Big \rbrace \bigcup \Big \lbrace 3 e^{i([0:5]\pi/3)} \Big \rbrace \bigcup \Big \lbrace 2\sqrt{3} e^{i(\pi/6+[0:5]\pi/3)} \Big \rbrace
\end{align}
Let $h_0$ denote the Rayleigh fast fade (with exponentially distributed channel power $|h_0|^2$ with unit mean)  between the central macrocell BS $C_0$ and its scheduled user $0$. Denoting the Euclidean norm by $\vectornorm{\cdot}$, the expression for the received SIR for macrocell user $0$ at position $\mathbf{r}$ is given as:
\begin{align}
\label{Eq:SIRMacro}
\mathrm{SIR}_c(\mathbf{r})&=\frac{\Theta_0}{\Psi_I(\mathbf{r})} {\lvert{h_0}\rvert}^2
\vectornorm{\mathbf{r}/R_c}^{-\alpha_c}\\
\textrm{where } \Psi_I(\mathbf{r})&= \sum_{k \in \cB} \vectornorm{\frac{\mathbf{r}-\mathbf{b}_k}{R_c}}^{-\alpha_c} {\lvert h_{0k}\rvert}^2 \Theta_{0k}
\end{align}
Here $\alpha_c$ represents the outdoor path loss exponent and $|h_{0k}|^2 \sim \mathrm{exp}(1)$ is the exponentially distributed channel power between interfering BS $C_k$ and the user of interest. The RV $\Theta_0$ [resp. $\Theta_{0k}$] is the lognormal shadowing between the central BS [resp. interfering BSs] and the desired user, which are distributed as $\mathrm{LN}(\zeta \mu_{c,\textrm{dB}},\zeta^2 \sigma_{c,\textrm{dB}}^2)$, where $\zeta=0.1 \ln{10}$ is a scaling constant. For analytical tractability in the reminder of the paper, our paper makes the following assumption regarding the distribution of a composite lognormal-exponential RV:
\begin{assumption}
\label{As:LNExp}
 The distribution of a composite lognormal-exponential RV $\Theta_k \abs{h_{0k}}^2$ is modeled as a lognormal distribution using Turkmani's  approximation \cite{Turkmani1992}:
\begin{align}
f_{\Psi_{i}}(x)&=\frac{1}{x \sigma_i\sqrt{2\pi}}\exp{\Biggl \lbrack \frac{-(\ln{x}-\mu_i)^2}{2 \sigma_i^2}\Biggr\rbrack} \\
\mu_i&=\zeta(\mu_{c,\textrm{dB}}-2.5), \sigma_i=\zeta \sqrt{\sigma_{c,\textrm{dB}}^2+5.57^2}
\end{align}
 \end{assumption}
\begin{assumption}
\label{As:SumLNExp}
For a fixed $\mathbf{r}$, using the moment generating function based technique described in Mehta \emph{et al.}\cite{Mehta2007}, the sum of $\lvert \cB \rvert$ independent, but not identically distributed lognormal RVs in the expression $\Psi_I(\mathbf{r})= \sum_{k \in \cB} (\vectornorm{\mathbf{r}-\mathbf{b_k}}/R_c)^{-\alpha_c} \Psi_k$ is approximated by a single lognormal RV with parameters $\mathbf{LN}(\mu_I(\mathbf{r}),\sigma_I^2(\mathbf{r}))$.
\end{assumption}
Using Assumptions \ref{As:LNExp}-\ref{As:SumLNExp} and \eqref{Eq:SIRMacro}, $\mathrm{SIR}_c(\mathbf{r})$ is distributed according to a lognormal RV $\mathrm{LN}(\mu_C(\mathbf{r}),\sigma_C^2(\mathbf{r})$ where $\mu_C(\mathbf{r})=\mu_S(\mathbf{r})-\mu_I(\mathbf{r})$ and $\sigma_C(\mathbf{r})=\sqrt{\sigma_S^2(\mathbf{r})+\sigma_I^2(\mathbf{r})}$. Then, the distribution of the SIR for a mobile tier 1 user at position $\mathbf{r}$ w.r.t the central BS is given by
\begin{align}
\label{Eq:SIRCDFMacro}
\Pr{[\mathrm{SIR}_c(\mathbf{r}) \leq \Gamma | \mathbf{r}]}&=
1-\mathrm{Q}\Biggl \lbrack \frac{\ln{(\Gamma \vectornorm{\mathbf{r}/R_c}^{\alpha_c})}-\mu_C}
{\sigma_C}
\Biggr \rbrack
\end{align}
where $\mathrm{Q}(x) \triangleq \frac{1}{\sqrt{2 \pi}}\int_{x}^{\infty}e^{-t^2/2} \mathrm{d}t$ is the complementary cumulative distribution function (CCDF) of a standard normal. Defining $a(\mathbf{r}) \triangleq \frac{\ln{\Gamma}-\mu_C(\mathbf{r})}{\sigma_C(\mathbf{r})}$ and $b \triangleq \frac{\alpha_c}{\sigma_C(\mathbf{r})}$, \eqref{Eq:SIRCDFMacro} simplifies to
\begin{align}
\label{Eq:SIRCDFMacro2}
\Pr{[\mathrm{SIR}_c(\mathbf{r}) \leq \Gamma | \mathbf{r}]}&=1-\mathrm{Q}[a(\mathbf{r})+b(\mathbf{r}) \ln{\vectornorm{\mathbf{r}/R_c}}]
\end{align}

\begin{figure}[htp]
  \begin{center}
    \subfigure[$T_c$ versus outdoor path-loss exponent $\alpha_c$]{\label{fig:MacroTPT} \includegraphics[width=3.25in]{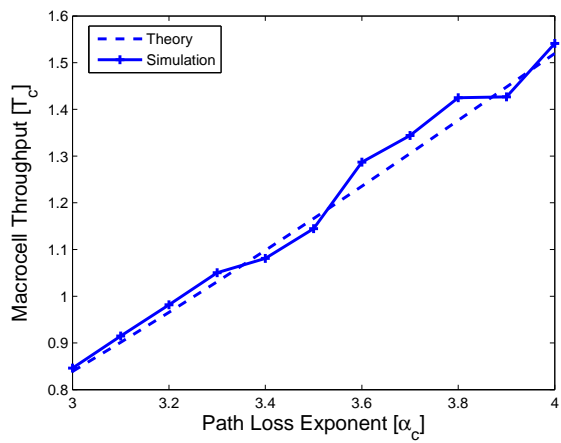}}
    \subfigure[$T_c$ with RR and PF Scheduling, $\alpha_c=4$]{\label{fig:MacroTPTRRVsPF} \includegraphics[width=3.25in]{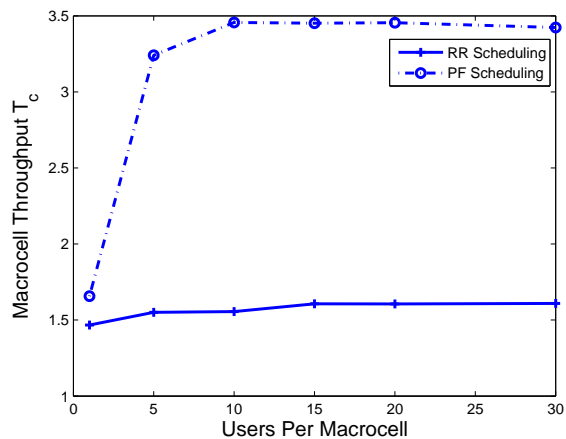}}
  \end{center}
  \caption{Spatially averaged macrocell subchannel throughput $T_c$ (b/s/Hz)}
  \label{fig:TPTMacro}
\end{figure}

Averaging \eqref{Eq:SIRCDFMacro2} over a hexagonal cell region is difficult. Alternatively, the spatially averaged CDF of $\mathrm{SIR}_c$ can be obtained approximately by considering an circular region of radius $\sqrt{\frac{\left|\cH \right|}{\pi}}$, which results in the same area as the cell site $\cH$.
To calculate the spatial throughput inside this circular region, the paper divides the region into $M$ non-overlapping annuli. For tractability, a simplifying assumption is that all users inside an annulus experience identical shadowing statistics (i.e. identical $\mu_C(\mathbf{r})$ and $\sigma_C(\mathbf{r})$). Denoting the distance of the user from $C_0$ by $\vectornorm{\mathbf{r}}=R$, the following lemma derives the expected spatial throughput by averaging $\mathrm{SIR}_c(R)$ inside a \emph{circular annulus} with inner radius $R_1$ and outer radius $R_2$.
\begin{lemma}
\label{Le:SIRCDFMacroAnn}
\emph{The spatially averaged SIR distribution inside a circular annulus with inner radius $R_1$ and outer radius $R_2$ is given as:}
\begin{align}
\label{Eq:SIRCDFMacroAnnular}
\mathbb{E}_{R}[\Pr{(\mathrm{SIR}_c \leq \Gamma | R_1 \leq R \leq R_2)}]&=1-\frac{1}{(R_{2}^2-R_{1}^2)}[R_{2}^2 C(a_{2},b)-R_{1}^2 C(a_{1},b)] \\
\label{Eq:SIRCDFMacroAnnC}
\textrm{where } C(a,b) & \triangleq \mathrm{Q}(a)+\exp \Biggl (\frac{2-2ab}{b^2}\Biggr)\mathrm{Q}\Biggl(\frac{2-ab}{b}\Biggr) \\
\label{Eq:SIRCDFMacroAnnAB}
a &\triangleq \frac{\ln{\Gamma}-\mu_C(R_2)}{\sigma_C(R_2)}, b \triangleq \frac{\alpha_c}{\sigma_C(R_2)} \\
\label{Eq:SIRCDFMacroAnnA1A2}
a_{2} &= a+b \ln{(R_{2}/R_c)}, a_{1}=a+b \ln{(R_{1}/R_c)},
\end{align}
\end{lemma}
\begin{proof}
See Appendix \ref{Proof:SIRCDFMacroAnn}.
\end{proof}
Lemma \ref{Le:SIRCDFMacroAnn} provides a simple method for estimating the cell-averaged macrocell throughput per subchannel. The probability that a user lies in an annulus with inner radius $R_{m-1}$ and outer radius $R_m$ ($1 \leq m \leq M$ with $R_0 = 0$) equals $\frac{\pi(R_m^2-R_{m-1}^2)}{|\cH|}$. We make use of assumptions \ref{As:LNExp} through \ref{As:SumLNExp} for computing the shadowing parameters $\sigma_C$ and $\mu_C$ at discrete locations $R_m, 1 \leq m \leq M$ where $R_M=\sqrt{\frac{|\cH|}{\pi}}$. The spatially averaged SIR distribution for a macrocell user is therefore approximated as follows:
\begin{align}
\label{Eq:EstSIRCDFMacro}
\Pr{(\mathrm{SIR}_c \leq \Gamma)} & =\mathbb{E}_R[\Pr{(\mathrm{SIR}_c(R) \leq \Gamma)}] \\
\label{Eq:EstSIRCDFMacroStp1}
                                & \approx \sum_{m=1}^{M} \mathbb{E}_R[\Pr{(\mathrm{SIR}_c \leq \Gamma | R_{m-1} \leq R \leq R_m)}] \cdot \frac{\pi(R_m^2-R_{m-1}^2)}{\left|\cH \right|} \\
\label{Eq:EstSIRCDFMacroStp2}
                                &= 1-\frac{\pi R_{1}^2} {|\cH|} C\left (a_1+b_1 \ln{\frac{R_1}{R_c}},b_1 \right) \notag \\
&-\sum_{m=2}^{M}\frac{\pi}{|\cH|} \left \lbrack R_{m}^2 C\left(a_m+b_m \ln{\frac{R_m}{R_c}},b_{m}\right)- R_{m-1}^2 C\left(a_m+b_m \ln{\frac{R_{m-1}}{R_c}},b_{m+1}\right)\right \rbrack \\
\textrm{where } a_m &\triangleq \frac{\ln{\Gamma}-\mu_C(R_m)}{\sigma_C(R_m)} \textrm{ and } b_m \triangleq \alpha_c/\sigma_C(R_m) \notag
\end{align}
where \eqref{Eq:EstSIRCDFMacroStp1} approximates \eqref{Eq:EstSIRCDFMacro} by spatially averaging $\mathrm{SIR}_c$ over $M$ different annulus. Equation \eqref{Eq:EstSIRCDFMacroStp2} is obtained by substituting \eqref{Eq:SIRCDFMacroAnnular} inside the conditional expectation in \eqref{Eq:EstSIRCDFMacroStp1} and the corresponding probability that the user lies in annulus $m, 1 \leq m \leq M$. Combining equations \eqref{Eq:MeanRate} and \eqref{Eq:EstSIRCDFMacro}, the average macrocell throughput $T_c$ in a given subchannel is expressed as:
\begin{align}
\label{Eq:TPTMacro}
T_c&=\sum_{l=1}^{L-1} l \cdot \mathbb{E}_{R}[\Pr(\Gamma_l \leq \mathrm{SIR}_c(R) < \Gamma_{l+1})]+L \cdot \mathbb{E}_{R}[\Pr(\mathrm{SIR}_c(R) \geq \Gamma_L)] \notag \\
   &=\sum_{l=1}^{L-1} l \cdot\left(\mathbb{E}_{R}[\Pr(\mathrm{SIR}_c(R) \leq \Gamma_{l+1})]-\mathbb{E}_{R}[\Pr(\mathrm{SIR}_c(R) < \Gamma_{l})]\right)+L \cdot \mathbb{E}_{R}[\Pr(\mathrm{SIR}_c(R) >\Gamma_L)]
\end{align}
Figure \ref{fig:MacroTPT} plots $T_c$ (in b/s/Hz) with RR scheduling as a function of the outdoor path-loss exponent $\alpha_c$ for the system parameters in Table \ref{Tbl:Sysprms}. The close agreement between theory and numerical simulations indicates that the theoretically obtained SIR distribution is an accurate approximation for practical throughput in a macrocellular environment.

\subsection{Macrocell Throughput: PF Scheduling}
In contrast to a RR scheduler, a PF scheduler enables macrocell users to compete for resources based on their requested rates normalized by their average throughput thus far. Consequently, the macrocell selects the user with the highest rate relative to their average rate. During the transmission interval $n$ in subchannel $m$, denote $R_k[m,n]$ as the requested rate for user $k, 1 \leq k \leq U_c$, located at position $\mathbf{r}_k$ w.r.t the central macrocell $C$. Let $\bar{R}_k[n]$ be the windowed mean throughput obtained by user $k$ over the $F_c$ frequency subchannels allocated for macrocell transmission. The PF scheduler selects the user $\tilde{k}$ whose current supportable rate is highest relative to their mean rate. The scheduling policy per subchannel $m$ with equal per-subchannel transmission powers (Assumption \ref{AS:TXPWR}) is described as:
\begin{align}
\tilde{k}(m,n)= \argmax_{1 \leq k \leq U_c} \frac{R_k[m,n]}{\bar{R}_k[n]}
\end{align}
Note that mobile user $k$ calculates $R_k[m,n]$ using \eqref{Eq:InstRate} and \eqref{Eq:SIRMacro} respectively. The windowed throughput per user prior to transmission interval $(n+1)$ is updated according to the rule:
\begin{align}
\bar{R}_k[n+1]=(1-\frac{1}{N})\bar{R}_k[n]+\frac{1}{N} \sum_{m=1}^{F_c} R_k[m,n]\mathbf{1}[k=\tilde{k}(m,n)], 1 \leq k \leq U_c
\end{align}
where $\mathbf{1}[\cdot]$ is the indicator function determining whether user $k$ is scheduled during transmission interval $n$ in frequency link or not.
The window size $N$ is a parameter that is selected considering the delay tolerance for each user. Choosing a smaller $N$ enables a given user to be scheduled more often, whereas choosing larger $N$ relaxes the fairness constraint and allows the scheduler to wait longer before scheduling a user.
By the strong law of large numbers, the average throughput per frequency subchannel for a given set of user positions is obtained from the sample average over a long duration and expressed as:
\begin{align}
\label{eq:MeanTPTPFGivenPos}
\mathbb{E}[\bar{R}(F_c,U_c)|\mathbf{r}_1,\cdots \mathbf{r}_{U_c}]= \lim_{n \rightarrow \infty}\frac{1}{n}\sum_{j=1}^{n} \sum_{m=1}^{F_c}\frac{R_{\tilde{k}}[m,j]}{F_c}, \tilde{k} \in \lbrace 1,2 \cdots U_c \rbrace
\end{align}
where the expectation on the left hand side is over the joint pdf of all channel gains between users and their serving and interfering BSs. The spatial averaged subchannel macrocell throughput is obtained by averaging \eqref{eq:MeanTPTPFGivenPos} w.r.t the joint pdf $f_{\mathbf{R}_1, \cdots \mathbf{R}_{U_c}}(\cdot)$ and given as:
\begin{align}
\label{eq:MeanTPTPF}
T_{c}(\rho,U_c)=\mathbb{E}_{\mathbf{R}_1,\cdots \mathbf{R}_{U_c}}[\mathbb{E}[\bar{R}(F_c,U_c)|\mathbf{R}_1=\mathbf{r}_1,\cdots \mathbf{R}_{U_c}=\mathbf{r}_{U_c}]]
\end{align}
Using \eqref{eq:MeanTPTPF} to compute $T_{c}(\rho,U_c)$ is analytically intractable. This paper resorts to numerical simulation to empirically estimate $T_c(\rho,U_c)$, which is used to derive the bandwidth partitioning. In the simulation, the number of subchannels is set as $F_c=1$ with a link bandwidth $W=15$ KHz and a PF window parameter $N=500$ OFDM symbols. Each mobile is moving at $v=13.34$ m/s ($30$ mph) and the per-link throughput \eqref{eq:MeanTPTPFGivenPos} is averaged over $500$ drops, with $8000$ trials/drop for modeling time-varying Rayleigh fading. The Rayleigh fading is held fixed over a duration $T_c=0.4/f_d$ where $f_d=\frac{v f_c}{3*10^8}$ is the doppler frequency at a carrier frequency $f_c=2$ GHz.  Figure \ref{fig:MacroTPTRRVsPF} compares the performance of PF (numerically evaluated) versus RR scheduling for different $U_c$. Exploiting channel variations through proportional fairness \emph{roughly doubles} the expected subchannel throughput.

\subsection{Femtocell Throughput}
Since femtocells are modeled as randomly placed on $\mathbb{R}^2$ according to a SPPP $\Phi_f$ with intensity $\lambda_f$, the interference experienced by a femtocell user depends on the distances of these interfering BSs relative to the user and their respective channel gains. Using F-ALOHA, the interfering femtocells form a marked SPPP $\Lambda_f \subseteq \Phi_f$ with intensity $\rho_f \lambda_f$. In a given frequency subchannel, the cochannel interference $I_{f,f}$ experienced by a user 0 within femtocell $F_0$ is given as:
\begin{align}
\label{Eq:FemtoCCI}
I_{f,f}=\sum \limits_{k \in \Lambda_f} A_{f} \Theta_{0k} \abs{h_{0k}}^2 |x_{0k}|^{-\alpha_{f}}
\end{align}
where $\Theta_{0k} \sim \textrm{LN}(\zeta \mu_{fo, \textrm{dB}},\zeta^2 \sigma^2_{fo,\textrm{dB}})$ represents the lognormal shadowing from femtocell $F_k$ to user 0 and $|h_{0k}|^2$ is the exponentially distributed channel power between interfering femtocell $F_k$ and user 0 inside $F_0$. Denoting the exponentially distributed channel power between $F_0$ and user 0 as $|h_0|^2$, the received SIR is given as:
\begin{align}
\label{Eq:SIRFemto}
\mathrm{SIR}_f&=\frac{B_{f} \Theta_{0} \abs{h_{0}}^2 |R_f|^{-\beta_{f}}}{\sum \limits_{k \in \Lambda_f} A_{f} \Theta_{0k} \abs{h_{0k}}^2 |x_{0k}|^{-\alpha_{f}}}
\end{align}
Here, user 0 is assumed to be on the edge of the home femtocell $F_0$ and $x_{0k}$ represents the locations of the interfering femtocells $F_k$ w.r.t user 0. The term $\Theta_0 \sim \textrm{LN}(\zeta \mu_{fi,\textrm{dB}}, \zeta^2 \sigma^2_{fi,\textrm{dB}})$ is the indoor lognormal shadowing, and $\Psi_0 \triangleq \Theta_{0} \abs{h_{0}}^2$ [resp. $\Psi_{0k} \triangleq \Theta_{0k} \abs{h_{0k}}^2$] are the effective channel gains from the desired [resp. interfering BSs]. The terms $\alpha_f$ [resp. $\beta_f$] represent the path-loss exponents resulting from interfering transmissions [resp. in-home transmissions] to the user of interest. A simple model is used to distinguish between the fixed losses arising from in-home and interfering transmissions: specifically, home users are insulated against interfering femtocell transmissions through double penetration losses arising from external wall partitions \cite{QComHomeNodeBProp2007}. Consequently, $A_f$ and $B_f$ (in dB) are related as
$A_{f,\textrm{dB}}=B_{f,\textrm{dB}}+2 P_{f,\textrm{dB}}$ where $P_f=\sqrt{\frac{A_f}{B_f}}$ is the wall penetration loss.

\begin{figure} [htp]
\begin{center}
   \includegraphics[width=4.0in]{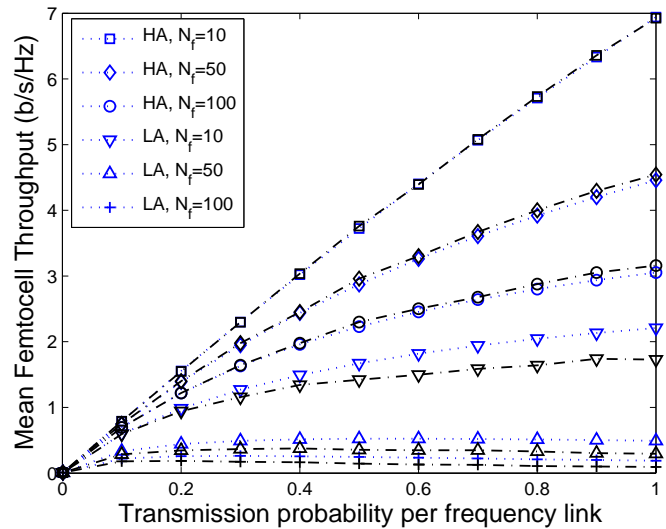}
   \caption{Theoretical and empirical throughput per femtocell $\rho_f T_f$ (b/s/Hz)}
   \label{fig:FemtoTPT}
   \end{center}
\end{figure}

Using Assumption \ref{As:LNExp}, the channel gain $\Theta_{0} \abs{h_{0}}^2$ is well approximated as a lognormal rv $\Psi_0 \sim \mathrm{LN}(\mu_{S},\sigma_{S}^2)$. Similarly, the channel gains $\Theta_{0k} \abs{h_{0k}}^2 \forall k$ are approximated as iid rv's distributed as $\Psi_I \sim \mathrm{LN}(\mu_I, \sigma_I^2)$.
Equation \eqref{Eq:SIRFemto} then simplifies to:
\begin{align}
\label{Eq:SIRFemto2}
\mathrm{SIR}_f&=\frac{\Psi_0 |R_f|^{-\beta_{f}}}{\sum \limits_{k \in \Lambda_f} P_f^2 \Psi_{0k} |x_{0k}|^{-\alpha_{f}}}
\end{align}
The closed form distribution of the Poisson SNP $I_{f,f}=\sum \limits_{k \in \Lambda_f} P_f^2 \Psi_{0k} |x_{0k}|^{-\alpha_{f}}$ is known only when $\alpha_f=4$\cite{Lowen1990}. However, tight lower bounds on $\Pr{(I_{f,f} > y)}$ are obtained by only considering femtocells whose interference \emph{individually} exceeds $y$. Using this idea, the following lemma, derived in Weber \emph{et. al.} \cite[Theorem 3]{Weber2007} provides an asymptotically tight lower bound on the tail distribution of $I_{f,f}$.
\begin{lemma}\cite[Theorem 3]{Weber2007}
\label{Le:FemtoCCDFLB}
\emph{With randomized transmissions and lacking power control, the lower bound on distribution of $I_{f,f}$ is given as:
\begin{align}
\label{Eq:FemtoCCDFLB}
\Pr(I_{f,f} > y) \geq 1-\exp{[-\pi \lambda_f \rho_f \mathbb{E}[{\Psi}_I^{\delta_f}]P_f^{2\delta_f}y^{-\delta_f}]}
\end{align}
where $\delta_f \triangleq \frac{2}{\alpha_f}$. When $\alpha_f=4$, $I_{f,f}$ is distributed as:
\begin{align}
\label{Eq:FemtoCCDFExact}
\Pr(I_{f,f} > y)=1-\mathrm{erfc}\Biggl (\frac{\pi^{3/2}\lambda_f \rho_f P_f \mathbb{E}[\Psi^{1/2}]}{2 \sqrt{y}}\Biggr)
\end{align}}
\end{lemma}
Lemma \ref{Le:FemtoCCDFLB} provides the relationship between the density $\lambda_f \rho_f$ of interfering femtocells in $\Lambda_f$ and the distribution of the CCI at a femtocell. For fixed $y$, as $\rho_f \rightarrow 0$, the tail probability $\Pr(I_{f,f}>y) \rightarrow 0$ in \eqref{Eq:FemtoCCDFLB} indicating that selecting fewer subchannels using F-ALOHA transmission provides greater resilience against persistent collisions from nearby femtocells. The distribution of the femtocell SIR in \eqref{Eq:SIRFemto} is obtained as:
\begin{align}
\Pr{(\mathrm{SIR}_f \leq \Gamma)} &=\Pr{\Biggl (\frac{\Psi_0 |R_f|^{-\beta_{f}}}{\sum \limits_{k \in \Lambda_f} P_f^2 \Psi_{0k} |x_{k}|^{-\alpha_{f}}} \leq \Gamma \Biggr)} \\
\label{Eq:FemtoSIRCDFS1}
    &=\mathbb{E}_{\Psi_0}\Biggl \lbrack \Pr{\Biggl (\sum \limits_{k \in \Lambda_f} P_f^2 \Psi_{0k} |x_{k}|^{-\alpha_{f}} \geq \frac{\psi_0 |R_f|^{-\beta_f}}{\Gamma}\Big \vert \Psi_0=\psi_0 \Biggr )} \Biggr \rbrack \\
\label{Eq:FemtoSIRCDFS2}
            & \geq 1-\mathbb{E}_{\Psi_0}\Biggl \lbrace \exp{\Biggl \lbrack -\pi \lambda_f \rho_f \mathbb{E}[{\Psi}_I^{\delta_f}] \Biggl (\frac{P_f^2 \Gamma}{\Psi_0 |R_f|^{-\beta_{f}}}\Biggr )^{\delta_f}\Biggr \rbrack}\Biggr \rbrace \\
\label{Eq:FemtoSIRCDF}
            &=1-\mathbb{E}_{\Psi_0}[\exp{(-\rho_f \kappa_f \Gamma^{\delta_f}\Psi_0^{-\delta_f})}] \\
\textrm{where, } \kappa_f & \triangleq \pi \lambda_f \mathbb{E}[{\Psi}_I^{\delta_f}] (P_f^2 |R_f|^{\beta_{f}})^{\delta_f} \notag
\end{align}
where \eqref{Eq:FemtoSIRCDFS1} and \eqref{Eq:FemtoSIRCDFS2} follow by conditioning on $\Psi_0$, assuming independence of $\Psi_0$ and $\Psi_{0k} \quad \forall k \in \Lambda_f$, and applying \eqref{Eq:FemtoCCDFLB}. Although it is not possible to obtain a closed form expression for the expectation in \eqref{Eq:FemtoSIRCDF}, the distribution of $\mathrm{SIR}_f$ can be calculated numerically. The mean subchannel throughput $T_f$ is calculated by combining \eqref{Eq:MeanRate} and \eqref{Eq:FemtoSIRCDF}:
\begin{align}
\label{Eq:TPTFemto}
T_f(\rho_f \lambda_f)&=\sum_{l=1}^{L-1} l \cdot \Pr(\Gamma_l \leq \mathrm{SIR}_f < \Gamma_{l+1})+L \cdot \Pr(\mathrm{SIR}_f \geq \Gamma_L) \\
\label{Eq:TPTFemto2}
                  & \approx \sum_{l=1}^{L-1} l \cdot \mathbb{E}_{\Psi_0}[\exp{(-\rho_f \kappa_f \Gamma_{l+1}^{\delta_f} \Psi_0^{-\delta_f})}-\exp{(-\rho_f \kappa_f \Gamma_l^{\delta_f}} \Psi_0^{-\delta_f})] \notag \\
 &+L \cdot \mathbb{E}_{\Psi_0}[\exp{(-\rho_f \kappa_f \Gamma_L^{\delta_f} \Psi_0^{-\delta_f})}]
\end{align}
The approximation in \eqref{Eq:TPTFemto} is because the right-hand side in \eqref{Eq:FemtoSIRCDF} is a lower bound on $\Pr(\mathrm{SIR}_f \leq \Gamma)$. Figure \ref{fig:FemtoTPT} plots the femtocell throughput $(1-\rho) \rho_f T_f$ (in b/s/Hz) assuming the entire bandwidth is allocated to femtocells ($\rho=0$). Black colored curves plot results of numerical simulations. Two cases are considered namely \begin{inparaenum}[(1)] \item high attenuation (marked ``HA" with $\alpha_f=4, P_{f,dB}=10$) and \item low attenuation (marked ``LA" with $\alpha_f=3.5, P_{f,dB}=2$) \end{inparaenum} from neighboring femtocells. Setting $\rho_f=1$ and assuming $N_f=50$ femtocells/cell site, the femtocell throughput falls from approximately $4.5$ b/s/Hz in a HA environment to nearly $0.5$ b/s/Hz in LA scenario, indicating the sensitivity of femtocell throughput to propagation from nearby femtocells.

To calculate the optimum $\rho_f$, we resort to maximizing the ASE per subchannel. This is analogous to answering the question: \emph{What fraction of subchannels should each femtocell access to maximize spatial reuse?} At this critical $\rho_f$, the F-ALOHA access by each femtocell is optimally traded off against neighboring femtocell interference in each subchannel. Mathematically, $\rho_f$ is the solution to the following optimization problem:
\begin{align}
\label{Eq:OptASEFemto}
\rho_f^{\ast} &= \lambda_f \arg \max_{0 < \theta \leq 1} \theta  T_f(\theta \lambda_f) \\
\mathrm{ASE}_f^{\ast}&=\rho_f^{\ast} \lambda_f T_f(\rho_f^{\ast} \lambda_f)
\end{align}
To justify \eqref{Eq:OptASEFemto}, observe that there are an average of $|\cH| \rho_f \lambda_f $ transmitting femtocells per subchannel. With F-ALOHA access of $0<\theta \leq 1$, each femtocell obtains an average subchannel throughput of $T_f(\theta)$, resulting in $\mathrm{ASE}_f$ equaling $\lambda_f \theta \cdot T_f(\theta \lambda_f)$. Alternatively, given any allocation $\rho$, \eqref{Eq:OptASEFemto} computes the F-ALOHA spectrum access $\rho_f$ which maximizes the mean \emph{overall} throughput $(1-\rho) \rho_f T_f(\rho_f \lambda_f)$ per femtocell.
\begin{remark}[\textbf{Boundedness of the ASE}]
\label{Re:OptASEFemto}
The ASE in \eqref{Eq:OptASEFemto} depends on the effective intensity $\lambda_f \theta$ of interfering femtocells per subchannel. With increasing $\lambda_f$, provided $\rho_f^{\ast}<1$, then the intensity of $\Lambda_f$ given as $\lambda_f \rho_f^{\ast}$ remains constant, implying that the optimal $\rho_f$ is a monotone decreasing function of $\lambda_f$. Consequently, if $\rho_f<1$ for a given $\lambda_f$, \emph{the maximum ASE per subchannel is fixed}. This also means that with increasing $\lambda_f$, the network-wide femtocell throughput equaling $|\cH| WF \cdot (1-\rho)\mathrm{ASE}_f^{\ast}$ grows \emph{linearly} with $(1-\rho)$.
\end{remark}

\begin{figure}[h]
  \begin{center}
    \subfigure[Femtocell ASE Vs F-ALOHA spectrum access]{\label{fig:ASEFemtoFALOHA}\includegraphics[width=4in]{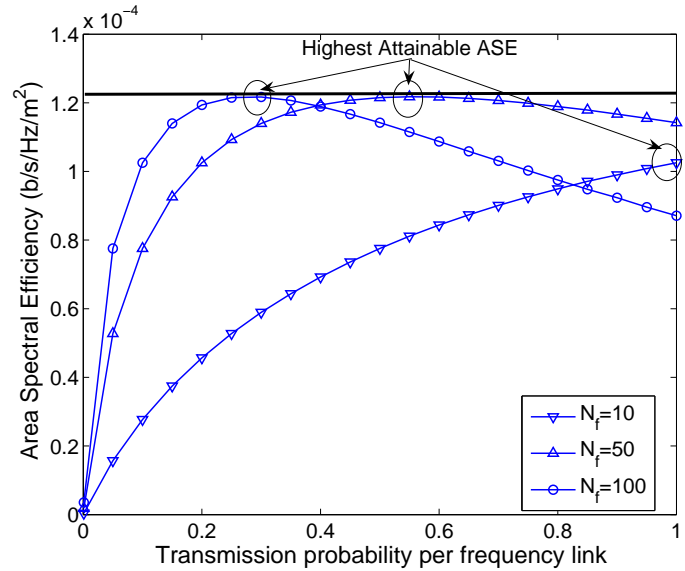}}
    \subfigure[Femtocell ASEs and subchannel throughputs]{\label{fig:FemtoSEASE} \includegraphics[width=4in]{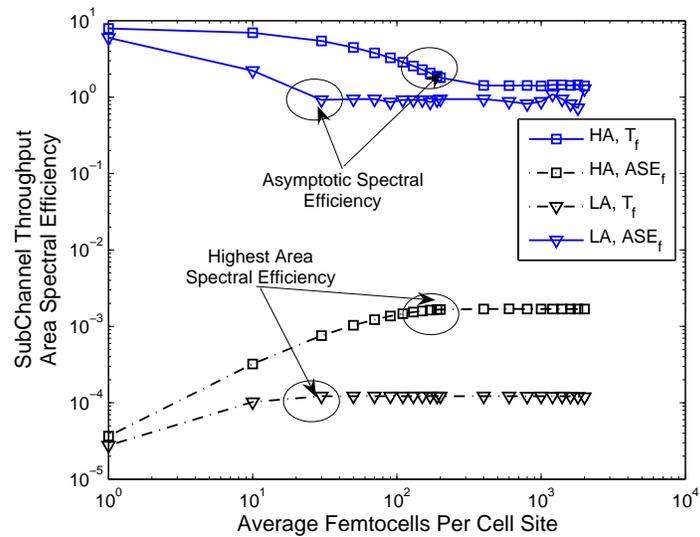}}
  \end{center}
  \caption{Femtocell Area Spectral Efficiencies $\frac{N_f \rho_f T_f}{|\cH|}$}
  \label{fig:ASEFemto}
\end{figure}

Fig. \ref{fig:ASEFemto} plots \eqref{Eq:OptASEFemto} for different $N_f$ with $\alpha_f=3.5$ and $P_{f,dB}=2$. In all cases, the highest ASE is fixed at nearly $0.000121 \textrm{ b/s/Hz/m}^2$ validating Remark \ref{Re:OptASEFemto}. With a low femtocell density ($N_f=10$), the best strategy is to access the entire spectrum from bandwidth partitioning. In a dense network ($N_f=100$), the ASE is maximized when each femtocell accesses approximately $30 \%$ of the available spectrum.
Further, in \eqref{Eq:OptASEFemto}, as long as $\rho_f^{\ast}=1$, each femtocell accesses the entire available spectrum $(1-\rho)$, consequently $T_f$ \emph{decreases} with addition of femtocells. However, if $\rho_f^{\ast}<1$, $T_f=\mathrm{ASE}_f/(\lambda_f \rho_f^{\ast})$ remains \emph{constant} with increasing $\lambda_f$ (Fig. \ref{fig:FemtoSEASE}). However, as $\lambda_f \to \infty$, since $\rho \in (0,1)$, the mean \emph{overall} throughput per femtocell approaches zero, as the following limit shows:
\begin{align}
\lim_{\lambda_f \rightarrow \infty} T_f (1-\rho) \rho_f \leq  \lim_{\lambda_f \rightarrow \infty} T_f \rho_f =0
\end{align}
One may explore the dependence of the mean overall femtocell throughput $T_f \rho_f (1-\rho)$ on the spectrum allocation $\rho$ and F-ALOHA access $\rho_f$. Equivalently: \emph{With increasing femtocell density $\lambda_f$, can increasing allocated spectrum $(1-\rho)$ to femtocells counterbalance decreasing F-ALOHA spectrum access $\rho_f$ to result in a higher mean femtocell throughput?}

This question is answered by the following condition:
Given an allocation $\rho_l$ at femtocell density $\lambda_f$, let $T_{f,l}$ and $\rho_{f,l}$ be the mean subchannel throughput and the optimal F-ALOHA access respectively. On increasing $\lambda_f$ by $\delta \lambda_f$ with allocation $\rho_h$, let the corresponding quantities equal $T_{f,h}$ and $\rho_{f,h}$. The femtocell network is defined as \emph{fully-utilized [resp. sub-utilized]} if a marginal increment in the femtocell density \emph{reduces [resp. improves]} the mean throughput per femtocell as given below:
\begin{align}
(1-\rho_l)\rho_{f,l}T_{f,l} &\gtrless (1-\rho_h)\rho_{f,h}T_{f,h} \notag \\
\Longleftrightarrow (1-\rho_l)\frac{\mathrm{ASE}_{f,l}}{\lambda_f} &\gtrless (1-\rho_h)\frac{\mathrm{ASE}_{f,h}}{\lambda_f+\delta \lambda_f} \notag \\
\label{Eq:FemtoUtil2}
\Longleftrightarrow \frac{1-\rho_h}{1-\rho_l} &\lessgtr \frac{\mathrm{ASE}_{f,l}}{\mathrm{ASE}_{f,h}} \cdot \frac{\lambda_f+\delta \lambda_f}{\lambda_f}=\frac{T_{f,l}}{T_{f,h}}\frac{\rho_{f,l}}{\rho_{f,h}}
\end{align}
Equation \eqref{Eq:FemtoUtil2} reflects the competing effects of increasing allocation $(1-\rho)$ and decreasing F-ALOHA access $\rho_f$ (or increasing $\lambda_f$) in determining the net femtocell throughput.

\section{Numerical Results}
\begin{figure}[h]
  \begin{center}
   \includegraphics[width=4in]{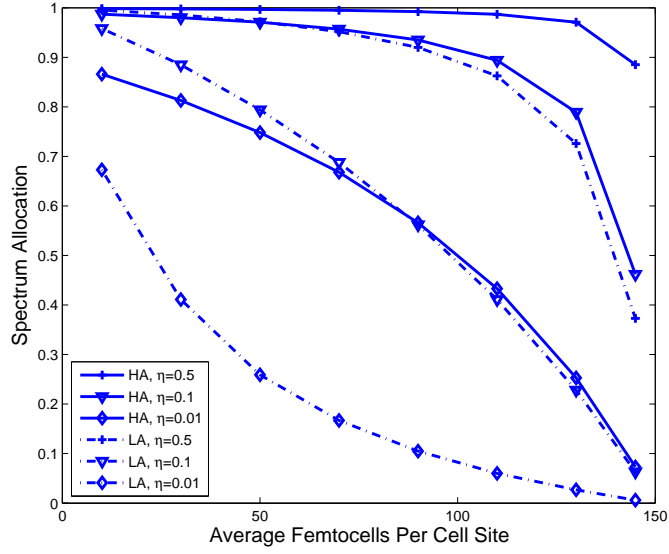}
   \caption{Optimal spectrum allocation $\rho$ for varying QoS parameter $\eta$}
   \label{fig:MacroFemtoBWPtgRR}
   \end{center}
\end{figure}

\begin{figure}[h]
\begin{center}
   \includegraphics[width=4in]{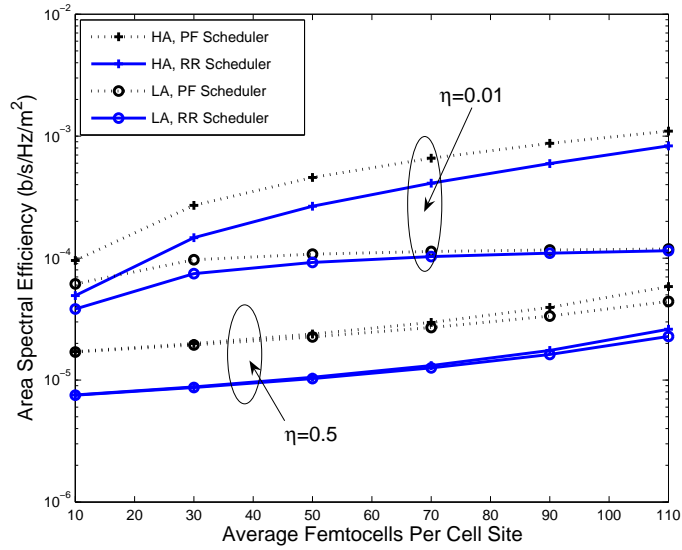}
   \caption{Area Spectral Efficiencies in a two-tier network for varying QoS parameter $\eta$}
   \label{fig:TwoTierASE}
   \end{center}
\end{figure}

\begin{figure} [htp]
\begin{center}
   \includegraphics[width=4in]{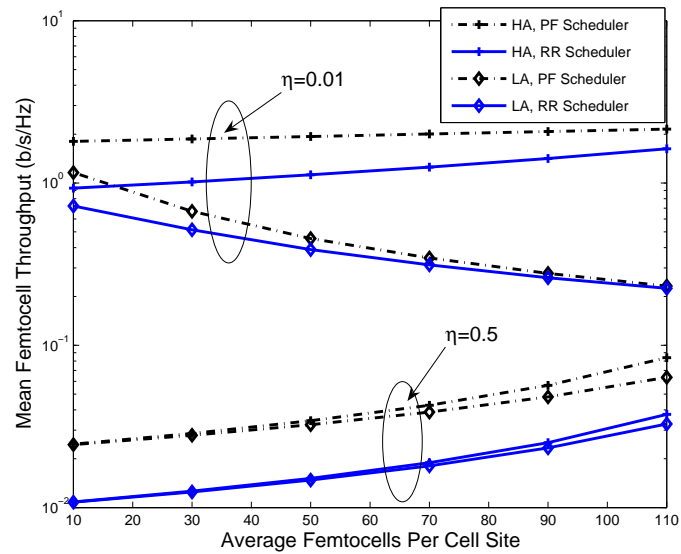}
   \caption{Average network-wide throughput $\rho_f T_f (1-\rho)$ (in b/s/Hz) provided by femtocells in their allocated spectrum  $(1-\rho)$}
    \label{fig:FemtoTPTBWPtg}
   \end{center}
\end{figure}

\begin{figure} [htp]
\begin{center}
   \includegraphics[width=4in]{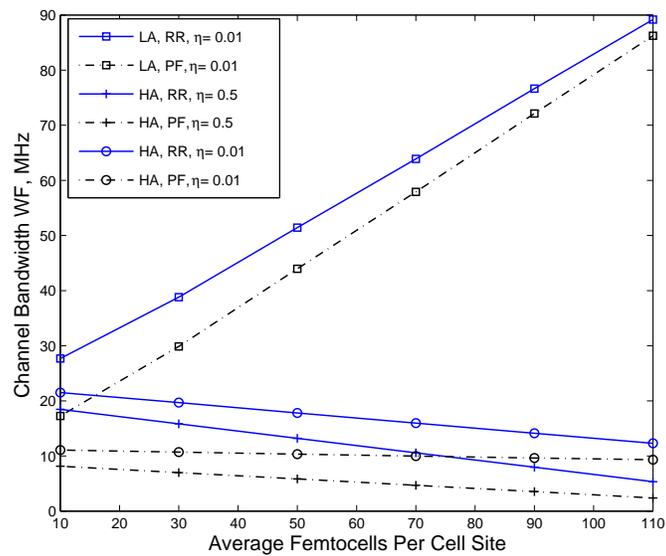}
   \caption{Required spectrum $WF$ meeting a target average data rate of $D_{c}=0.1$ Mbps for each macrocell user, given Round-Robin and Proportional-Fair scheduling at the macrocell. }
    \label{fig:ReqChannelBW}
   \end{center}
\end{figure}

Results are presented in Figs. \ref{fig:MacroFemtoBWPtgRR} through \ref{fig:ReqChannelBW} with the system parameters in Table \ref{Tbl:Sysprms}.  The number of users in each tier is controlled by varying $N_f$.  To model varying data-rate requirements inside femtocells relative to the central macrocell, QoS values of $\eta=0.5$ (equal per-user throughputs in each tier) and $\eta=0.01$ (favoring $100$x higher throughput/femtocell user relative to macrocell users) are considered. Two propagation scenarios are presented namely \begin{inparaenum}[1)]\item \emph{High Attenuation (HA)} of neighboring femtocell transmissions--modeling low interference--with parameters $\alpha_f=4$ and $P_{f,dB}=10$ and \item \emph{Low Attenuation (LA)} scenario by setting $\alpha_f=3.5$ and $P_{f,dB}=2$. \end{inparaenum}

Fig. \ref{fig:MacroFemtoBWPtgRR} shows the allocation using \eqref{Eq:ASEOptRho} with RR scheduling per-tier. With equal average per-user throughputs ($\eta=0.5$), nearly $90 \%$ of the overall bandwidth is assigned to the macrocell. The central macrocell serves a higher number of users, who experience relatively poor reception. Equalizing per-user throughputs consequently requires a significantly higher allocation to the macrocell. As $\eta$ decreases, femtocells require more spectrum for providing greater indoor capacity; eg. in a LA scenario with $\eta=0.01$ and $N_f=50$ femtocells/cell site, nearly $70 \%$ of spectrum is allocated to femtocells.

Fig. \ref{fig:TwoTierASE} plots the ASEs of the two-tier network using \eqref{Eq:NetworkASE}. In a LA scenario with $\eta=0.01$, the high degree of co-channel interference results in the ASE maximized with fewer than $N_f=50$ femtocells.  Following Remark \ref{Re:OptASEFemto}, this indicates that adding more femtocells does not provide additional spatial reuse. In all other cases, the ASEs monotonically increase with $N_f$ indicating increasing spatial reuse with addition of femtocells. To show benefits of opportunistic scheduling, a PF scheduler provides nearly $2.3$x [resp. $1.35$x] ASE gains relative to a RR scheduler in a HA scenario with QoS parameter $\eta=0.5$ [resp. $\eta=0.01$] and $N_f=110$ femtocells/cell site.

Fig. \ref{fig:FemtoTPTBWPtg} plots the expected throughput per femtocell $(1-\rho)\rho_f T_f$ as a function of $N_f$ and $\eta$. For $\eta=0.5$, the throughputs monotonically increase with $N_f$ indicating that increasing spectrum allocation $(1-\rho)$ counteracts the effects of decreasing $\rho_f T_f$; in effect, the femtocell network is sub-utilized. With $\eta=0.01$ in a LA environment however, the femtocell throughputs decrease with increasing $N_f$, indicating that the femtocell network is fully-utilized.

Fig. \ref{fig:ReqChannelBW} plots the minimum required spectrum $WF$, which satisfies a target average data rate of $D_{c}=0.1$ Mbps for each macrocell user. For each femtocell, corresponding to QoS parameter values $\eta=0.5$ and $\eta=0.01$, we consider target average data rates $D_f = D_c (1-\eta)/\eta$ equaling $0.1$ and $10$ Mbps/user.  Since Proposition \ref{Pr:TwoTierOptASE} ensures that the QoS constraint in \eqref{Eq:TwoTierOptASE} is binding, the required spectrum $WF$ is given as:
\begin{align}
WF=\frac{U_c}{\rho T_c}D_{c}= \frac{U_f}{(1-\rho)\rho_f T_f}D_{f}
\end{align}

Two key observations are: First, a channel aware macrocell provides significant savings in the spectrum necessary to meet $D_f$ and $D_c$; eg. with $\eta=0.01$ and $N_f=50$ femtocells/cell site in a HA scenario, the spectrum reduction is nearly $50\%$ ($10$ MHz). Next, spatial reuse and spectrum requirements with the addition of femtocells are markedly different depending on attenuation from neighboring femtocells. For example, in a LA [resp. HA] scenario, the spectrum requirement $WF$ increases [resp. decreases] with increasing hotspot density indicating the femtocell network is fully-utilized [resp. sub-utilized] with the per-tier spectrum allocation in \eqref{Eq:ASEOptRho}.

\section{Conclusions}
This paper has proposed a decentralized spectrum allocation strategy as an alternative to centralized/coordinated frequency assignment in a two-tier network. The proposed allocation depends on the per-tier throughputs, the loading of users in each tier and the QoS requirements, accounting for co-channel interference and path-losses. With a randomized spectrum access strategy, femtocells should access a decreasing fraction of their allocated spectrum with increasing femtocell density, in order to maximize spatial reuse. Spatial reuse benefits derived from channel aware macrocell scheduling result in nearly $50\%$ spectrum reduction for meeting target per-tier data rates. In a low interference scenario where addition of hotspots provides increased spatial reuse, the spectrum requirement is unchanged up to $110$ femtocells/cell site. On the other hand, the limited spatial reuse in high interference scenarios necessitates increasing spectrum with addition of femtocells. These insights provide guidelines on performance of decentralized spectrum allocation in the two-tier networks.

\appendices
\section{}
\label{Proof:SIRCDFMacroAnn}
Inside a circular annulus of small width, one can assume that a user experiences identical shadowing statistics from interfering BSs. For convenience, the parameters $a$ and $b$ are chosen when the user is at the outer edge ($R=R_2$) of the annulus, as shown in equation \eqref{Eq:SIRCDFMacroAnnAB}. Given $R_1 \leq R \leq R_2$ with the pdf function $f_{R}(r|R_1  \leq R \leq R_2)=\frac{2r}{R_2^2-R_1^2}$, the spatially averaged outage probability is obtained as:
\begin{align}
\label{Eq:SIRCDFMacroAnn}
\mathbb{E}_{R}[\Pr{(\mathrm{SIR}_c \leq \Gamma |R_1 \leq R \leq R_2)}]&= \notag \\
& 1-\mathbb{E}_{R}[\mathrm{Q}(a+b \ln{R/R_c}) \mid R_1 \leq R \leq R_2] \\
\label{Eq:SIRCDFMacroAnnStp1}
&=1-\frac{2}{R_2^2-R_1^2}\int_{R_1}^{R_2} \mathrm{Q}[a+b\ln{r/R_c}] r \mathrm{d}r \\
\label{Eq:SIRCDFMacroAnnStp2}
&=1-\frac{1}{R_2^2-R_1^2}\Biggl \lbrace
           R_2^2 \int_{0}^{R_2} \frac{2}{R_2^2} \mathrm{Q}[a_2+b\ln{r/R_2}] r \mathrm{d}r \notag \\
        &- R_1^2 \int_{0}^{R_1} \frac{2}{R_1^2} \mathrm{Q}[a_1+b\ln{r/R_1}] r \mathrm{d}r \Biggr \rbrace
\end{align}
where \eqref{Eq:SIRCDFMacroAnnStp2} is obtained by substituting $a_2 \triangleq a+b \ln {R_2/R_c}$ and $a_1 \triangleq a+b \ln{R_1/R_c}$ in \eqref{Eq:SIRCDFMacroAnnStp1}. Combining the definitions in \eqref{Eq:SIRCDFMacroAnnC} and \eqref{Eq:SIRCDFMacroAnnA1A2} with the identity \cite[Pg. 55]{Goldsmith}:
\begin{align}
C(a,b)=\frac{2}{R^2}\int_{0}^{R} \mathrm{Q}(a+b\ln{r/R}) r \mathrm{d}r
\end{align}
and plugging into \eqref{Eq:SIRCDFMacroAnnStp2}, the result follows.

\begin{table}[ht]
\caption{System Parameters} \label{Tbl:Sysprms} \centering
\begin{tabular}{ c| c | c}
    \hline
    \textbf{Symbol} & \textbf{Description} & \textbf{Value} \\ \hline
     $R_c,R_f$  & Macrocell/Femtocell Radius     & $288$ m, $40$ m\\
     $U$    &  Total users per cell site & 300 \\
     $U_f$      & Users per femtocell & $2$ \\
    $P_{f,\textrm{dB}}$ & Wall penetration loss & $2$ dB, $10$ dB \\
    $G, L$   & Shannon Gap, Modulation Levels   & $3$ dB, 8 \\
     $\alpha_c$ & Path-loss exponent (Macrocell Outdoor) &  $4$ \\
    $\alpha_f$ & Path-loss exponent (Femtocell to Femtocell) &  $3.5,4$ \\
    $\beta_f$ &  Path-loss exponent (Inside Home Femtocell) &  $3$ \\
    $\sigma_{c,\textrm{dB}},\sigma_{fi,\textrm{dB}},\sigma_{fo,\textrm{dB}}$ & Lognormal Shadow Parameters & $8$ dB, $4$ dB, $12$ dB \\
    \hline
\end{tabular}
\end{table}

\end{document}